\documentclass[twoside]{article}
\usepackage{fleqn,espcrc2,epsfig,amsmath,amssymb}

\title{Dirac eigenvalues and eigenvectors at finite
    temperature\thanks{Presented by TW at Lattice 2000, Bangalore,
    India.}} 
\author{M. G\"ockeler\address[Regensburg]{Institut f\"ur Theoretische
    Physik, Universit\"at Regensburg, D-93040 Regensburg, Germany},
  H.Hehl\addressmark[Regensburg],
  P.E.L. Rakow\addressmark[Regensburg], 
  A. Sch\"afer\addressmark[Regensburg],
  W. S\"oldner\addressmark[Regensburg], 
  and T. Wettig\address[Yale]{Center for Theoretical Physics, Yale 
    University, New Haven, CT 06520-8120, USA and\\
    RIKEN BNL Research Center, Brookhaven National Laboratory, Upton,
    NY 11973-5000, USA}}
       
\begin{document}

\begin{abstract}
  We investigate the eigenvalues and eigenvectors of the staggered
  Dirac operator in the vicinity of the chiral phase transition of
  quenched SU(3) lattice gauge theory.  We consider both the global
  features of the spectrum and the local correlations.  In the
  chirally symmetric phase, the local correlations in the bulk of the
  spectrum are still described by random matrix theory, and we
  investigate the dependence of the bulk Thouless energy on the
  simulation parameters.  At and above the critical point, the
  properties of the low-lying Dirac eigenvalues depend on the
  $Z_3$-phase of the Polyakov loop.  In the real phase, they are no
  longer described by chiral random matrix theory.  We also
  investigate the localization properties of the Dirac eigenvectors in
  the different $Z_3$-phases.  
\end{abstract}
\maketitle

\vspace{-3mm}

\section{INTRODUCTION}

The theoretical understanding of the Dirac spectrum has improved
considerably in the past few years.  Using a variety of methods such
as finite volume partition functions \cite{Leut92}, partially quenched
chiral perturbation theory \cite{Bern92}, and chiral random matrix
theory (RMT) \cite{Shur93}, it has been shown that in the phase in
which chiral symmetry is spontaneously broken, the distribution and
the correlations of the low-lying Dirac eigenvalues are described by
relatively simple universal functions.  This description is valid in a
regime in which the zero-momentum modes dominate the effective
Lagrangian.  The energy scale which limits this regime is known as the
Thouless energy.  For a review, we refer to Ref.~\cite{review}.

Recently, two studies \cite{Farc99b,Damg00} have appeared in which the
Dirac spectrum was investigated in this context for temperatures $T$
close to the critical temperature $T_c$ of the chiral phase
transition.  This is an interesting problem, since the above-mentioned
approach only works in the broken phase, and one would like to find
out what happens to the universal features as one crosses $T_c$.  The
present study addresses these and related questions.  In addition, we
also investigate the properties of the Dirac eigenvectors.

\section{$\mathbf{Z_3}$ ENSEMBLES}

\begin{figure}[!t]
  \centerline{\epsfig{figure=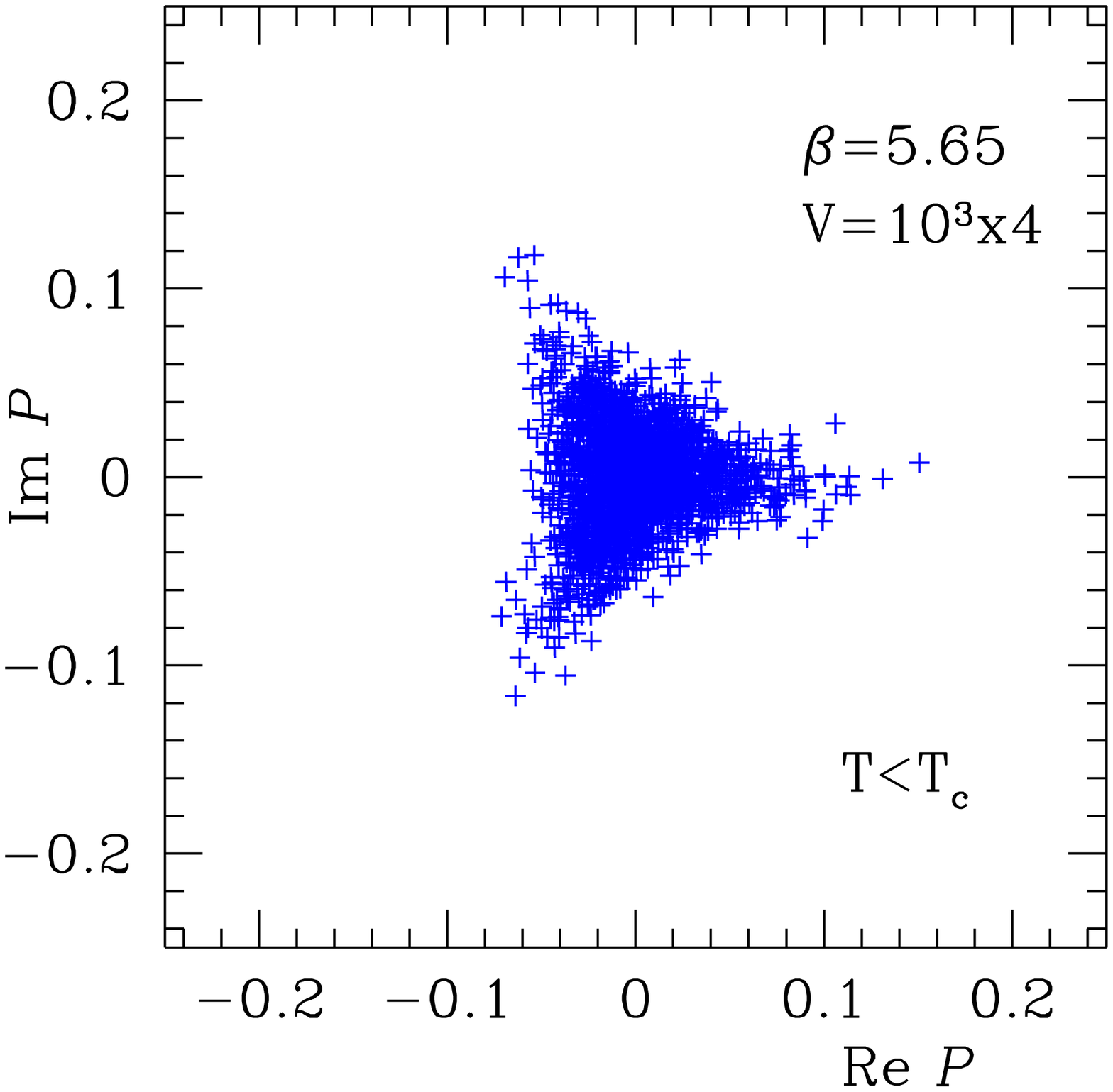,width=35.4mm}
    \hspace{3mm}\epsfig{figure=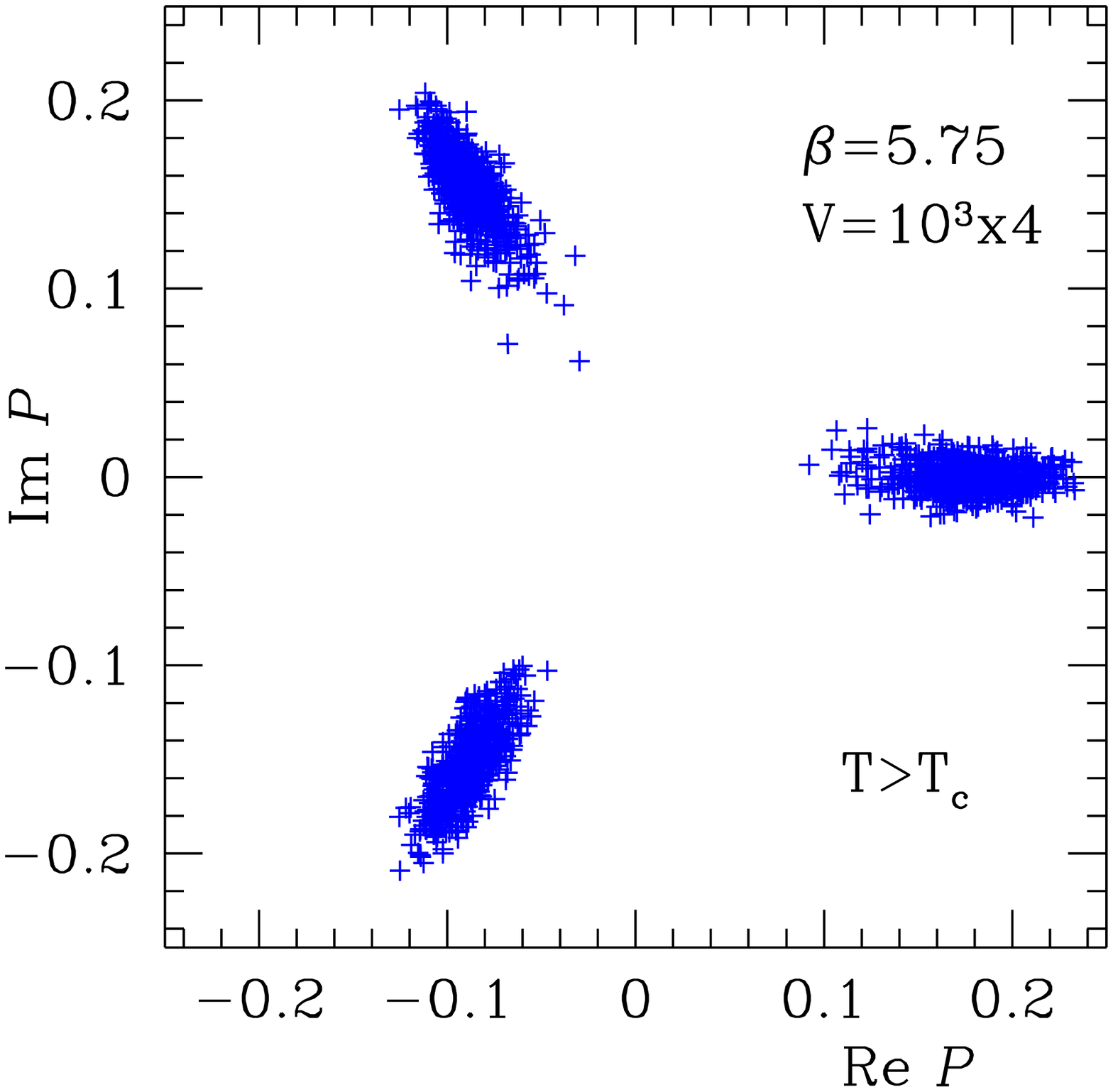,width=35.4mm}}
    \vspace*{-8mm}
    \caption{Scatter plot of the Polyakov loop in the complex plane in
      the confinement (left) and deconfinement (right) phase.}
    \label{fig:phases}
    \vspace*{-5mm}
\end{figure}

We are working in the quenched approximation using the staggered
discretization of the Dirac operator.  In a study of chiral symmetry
restoration in the quenched approximation, an interesting observation
was made in Ref.~\cite{Chan95}.  The gauge action has a $Z_3$ symmetry
(for $N_c=3$ colors) which is broken in the deconfinement phase.  As a
result, the phases of the Polyakov loop $P$ cluster around the
elements of $Z_3$ in the complex plane, and one can divide the total
ensemble of gauge field configurations into three classes with
arg($P)=0,\pm2\pi/3$.  An example is shown in Fig.~\ref{fig:phases}.
It was found \cite{Chan95} that the chiral condensate computed from
the class of configurations with arg($P)=0$ vanishes above $T_c$ as
expected.  However, for arg($P)=\pm2\pi/3$ it remains nonzero in a
certain range of $T$ above $T_c$.  This behavior can be understood
qualitatively in NJL-type models \cite{Meis95,Chan95b} and in RMT
\cite{mishat}.  The point is that the boundary conditions of the Dirac
operator are not invariant under $Z_3$ transformations, and for
arg($P)=\pm2\pi/3$, the new boundary conditions lead to a shift of the
Dirac eigenvalues to lower values \cite{mishat}.  Using the
Banks-Casher relation \cite{Bank80}, this implies a nonzero chiral
condensate.  

In the following analysis, we therefore separate our configurations
into two ensembles, those with arg($P)=0$ (ensemble E1) and those with
arg($P)=\pm2\pi/3$ (ensemble E2), respectively.  This separation can
be done unambiguously.  For the purpose of the present analysis, the
two classes arg($P)=\pm2\pi/3$ are equivalent and can be combined in
E2.  In full QCD, the fermion determinant suppresses the E2
configurations.  (In Ref.~\cite{Damg00}, the E2-configurations were
$Z_3$-rotated before the Dirac operator was diagonalized so that
arg($P)=0$ in their analysis \cite{Urs}.)

Strictly speaking, we should distinguish three critical temperatures,
$T_d$ for the deconfinement phase transition, and $T_{c1}$ and
$T_{c2}$ for the chiral phase transitions of E1 and E2, respectively.
Here, we assume that $T_d\approx T_{c1}$.  Since we are mainly
interested in the region $T\approx T_{c1}$, we write $T_c$ instead of
$T_{c1}$ in the following.

\section{DIRAC SPECTRUM}

\begin{figure}[!b]
  \vspace*{-2mm}
  \centerline{\epsfig{figure=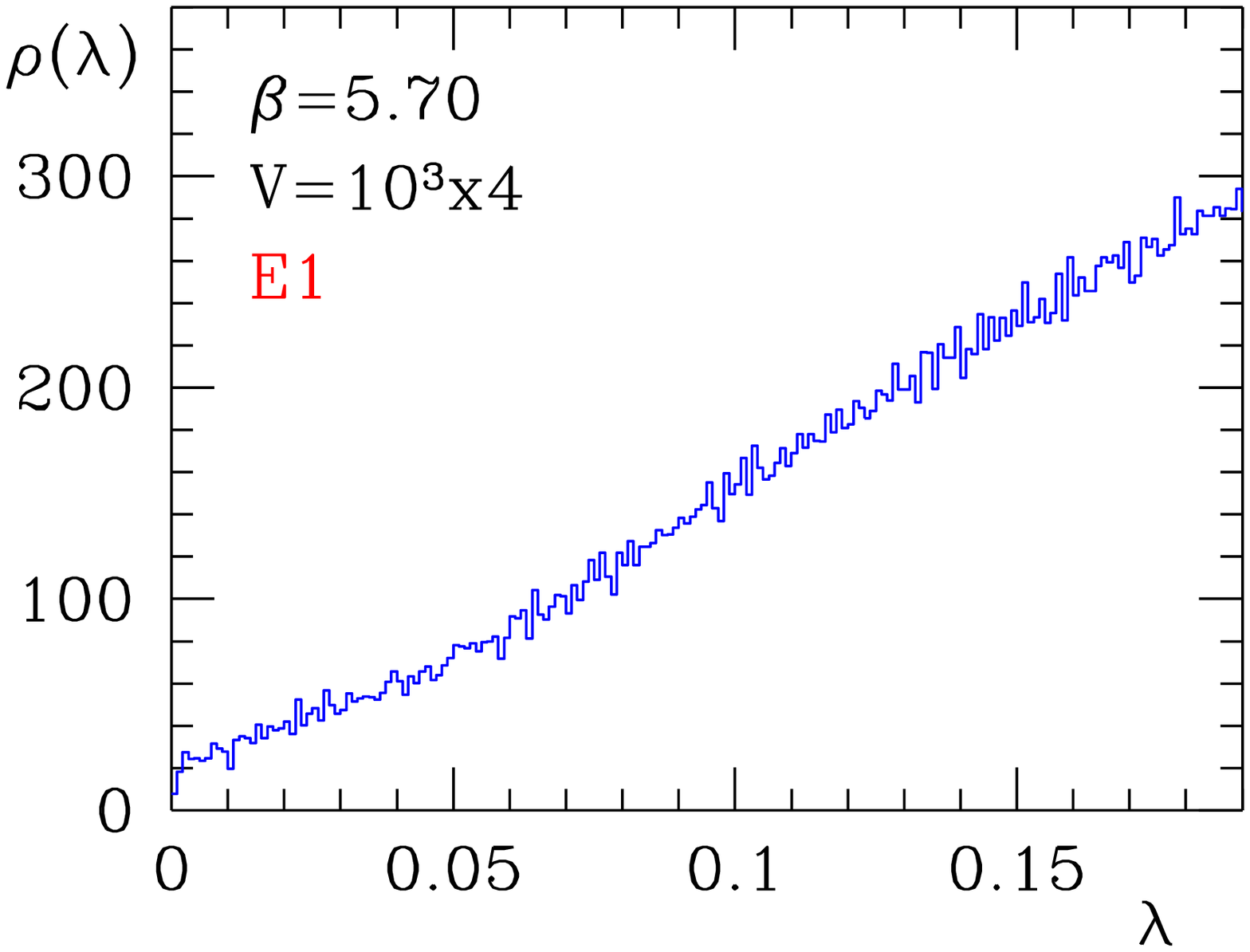,width=35.4mm}
    \hspace{3mm}\epsfig{figure=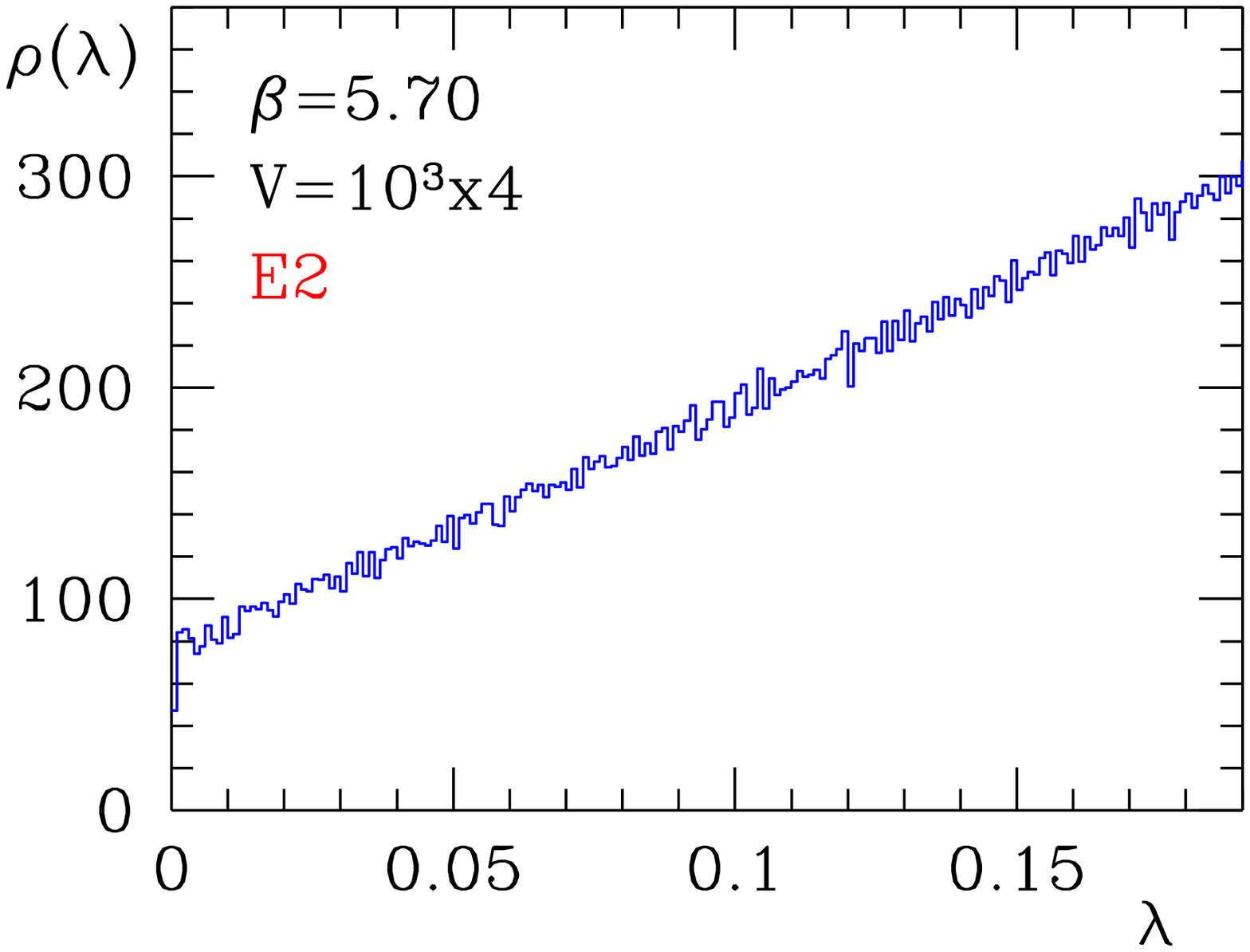,width=35.4mm}}
  \vspace*{-8mm}
  \caption{Global spectral density of the Dirac operator at
    $T\gtrsim T_c$ for the two different ensembles.}
  \label{fig:global}
\end{figure}

We have worked on $N_s^3\times N_t$ lattices with $N_t$=4 and 6 for
which $\beta_c(N_s\to\infty)$=5.6925 and 5.8941, respectively
\cite{Karsch}.  An example for the global spectral density of the
Dirac operator near zero for $T\gtrsim T_c$ is shown in
Fig.~\ref{fig:global}.
For the E1-ensemble, we find $\rho(0)\approx0$ which, according to the
Banks-Casher relation, implies that chiral symmetry is restored.  On
the other hand, for E2 we observe that $\rho(0)\ne0$ which implies a
nonzero chiral condensate, in agreement with \cite{Chan95}.

We therefore expect the distribution of the low-lying Dirac
eigenvalues to be different in the two ensembles.  Here, we
concentrate on the smallest positive eigenvalue,
$\lambda_{\text{min}}$.  Depending on whether or not
$\langle\bar\psi\psi\rangle\ne0$, the expectation value
$\langle\lambda_{\text{min}}\rangle$ scales as follows:
\begin{align}
  T<T_c:&\quad \lambda_{\text{min}}\sim V^{-1}\:,\notag\\
  T>T_c:&\quad \lambda_{\text{min}}\sim V^0\:,\\
  T=T_c:&\quad \lambda_{\text{min}}\sim V^{-\delta/(\delta+1)}
  \:,\notag
\end{align}
where $\delta$ is one of the universal critical exponents of a second
order phase transition.  For $T<T_c$, the distribution of
$\lambda_{\text{min}}$ is described by the RMT result,
$P(\lambda_{\text{min}})= (c^2\lambda_{\text{min}}/2)
\exp(-(c\lambda_{\text{min}})^2/4)$ with
$c=V|\langle\bar\psi\psi\rangle|$.  At $T=T_c$, RMT predictions for
$P(\lambda_{\text{min}})$ \cite{Akem98} are model dependent.

\begin{figure}[!t]
  \begin{center}
    \epsfig{figure=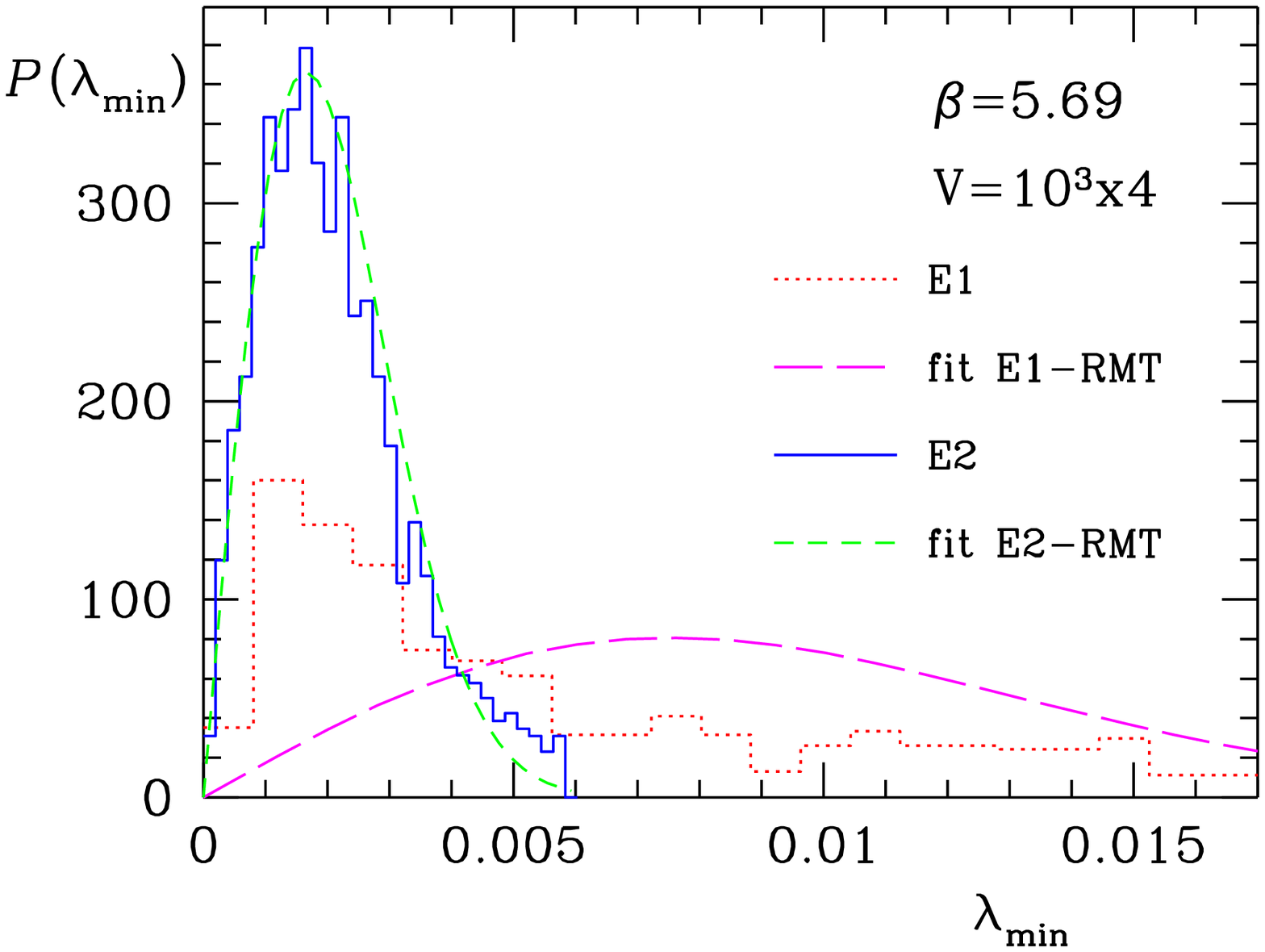,width=68mm}\\[3mm]
    \epsfig{figure=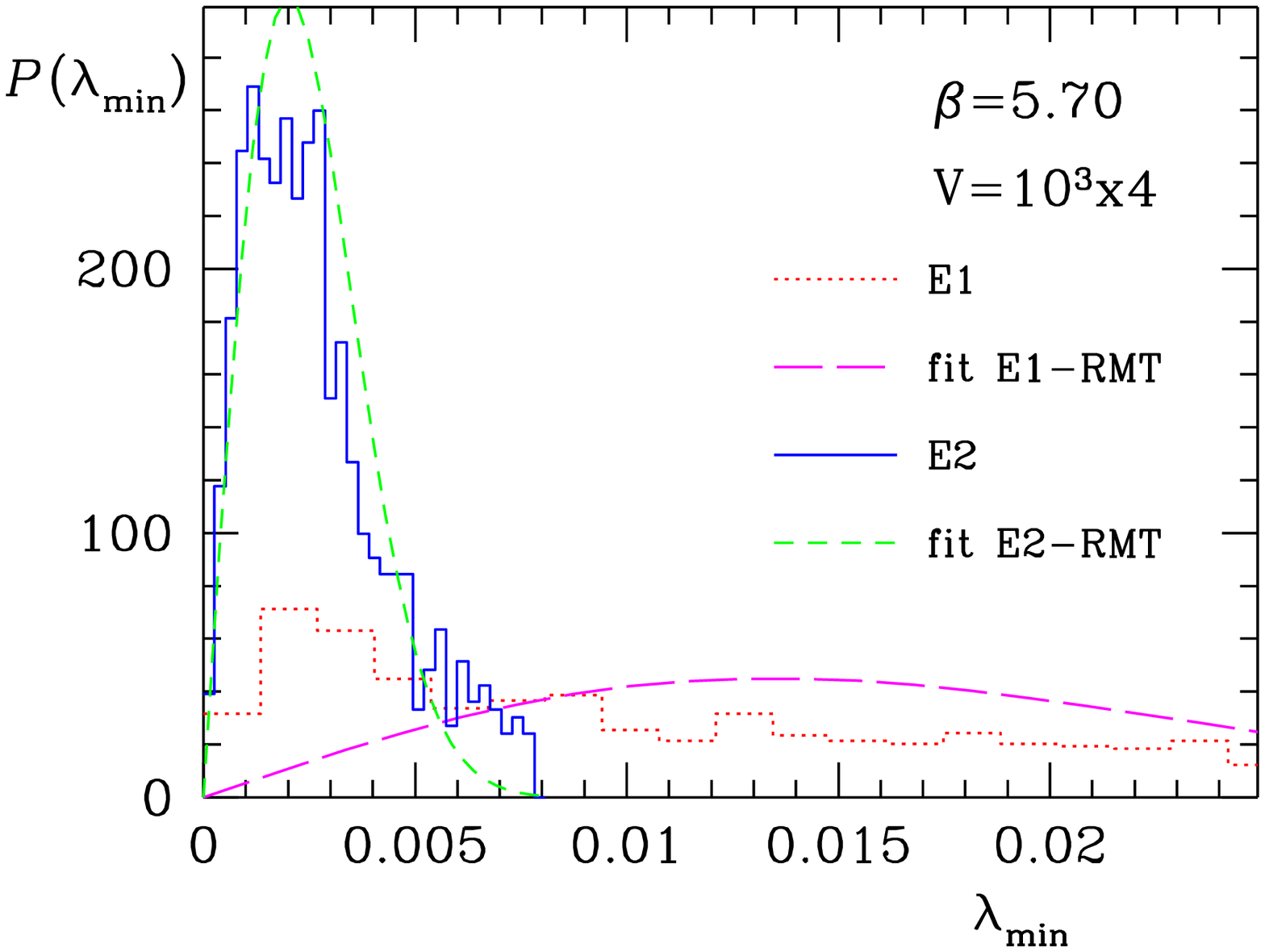,width=68mm}\\[-8mm]
    \caption{Distribution of the smallest Dirac eigenvalue (from 2000
      configurations) in the two ensembles for $T\approx T_c$ (top)
      and $T\gtrsim T_c$ (bottom).}
    \label{fig:smallest}
    \vspace*{-8mm}
  \end{center}
\end{figure}
Our results for $P(\lambda_{\text{min}})$, along with fits to the RMT
prediction in the broken phase, are shown in Fig.~\ref{fig:smallest}.
As expected, the results for the two ensembles E1 and E2 are very
different.  The top figure corresponds to $T\approx T_c$, and the
bottom figure to $T$ slightly above $T_c$.  In both figures,
$\langle\lambda_{\text{min}}\rangle$ is much larger in E1 than in E2.
This reflects the fact that in the symmetric phase, small eigenvalues
are suppressed.  Also, $P(\lambda_{\text{min}})$ in E1 is clearly not
described by the RMT prediction (which is valid for $T<T_c$).  For E2
at $T\approx T_c$, however, $P(\lambda_{\text{min}})$ is still very
well described by RMT, consistent with the fact that chiral symmetry
is still broken for this ensemble.  For $T>T_c$ the agreement becomes
worse. 

Other quantities such as the microscopic density could be analyzed in
exactly the same way, and the results and conclusions will be similar.

\section{THOULESS ENERGY FOR $\mathbf{T>T_c}$}

\begin{figure}[!t]
  \centerline{\epsfig{figure=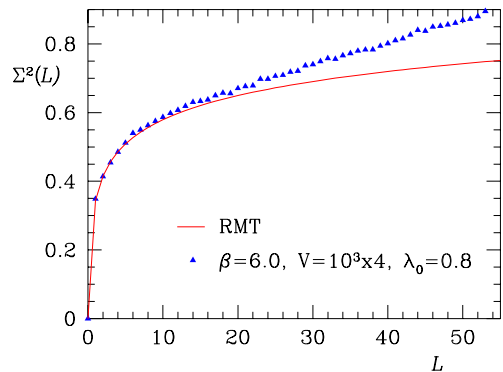,width=65mm}}
  \vspace*{-9mm}
    \caption{Number variance of the Dirac eigenvalues in the bulk of
      the spectrum.  $\lambda_0$ is the starting point of the interval
      of length $L$.}
    \label{fig:sigma2}
  \vspace*{-6mm}
\end{figure}

The Thouless energy \cite{Thou74} is the limiting energy above which
the universal description of the Dirac spectrum in terms of RMT is no
longer valid.  One has to distinguish between the Thouless energy at
the hard edge and in the bulk of the spectrum.  The Thouless energy at
the hard edge is very well understood, both theoretically
\cite{Osbo98} and on the lattice \cite{Berb98}.  For $T>T_c$ the hard
edge is no longer described by RMT so that the concept of a Thouless
energy no longer exists.  However, in the bulk the local spectral
correlations are still given by RMT, and it is interesting to study
the bulk Thouless energy above $T_c$.  A convenient measure of the
bulk spectral correlations is the number variance defined by
\begin{equation}
  \Sigma^2(L)=\big\langle\big(n(L)-\langle n(L)\rangle\big)^2\big\rangle\:,
\end{equation}
where $n(L)$ is the number of levels in an interval of length $L$
after the spectrum has been unfolded.  There are several questions
related to how the spectrum should be unfolded, see Ref.~\cite{Guhr99}
for a comprehensive discussion.  We have used ensemble averaging to
construct the average spectral density used in the unfolding
procedure.  The number variance was also computed by ensemble
averaging.  A typical example for $\Sigma^2(L)$, averaged over
600 independent configurations, is shown in Fig.~\ref{fig:sigma2},
along with the parameter-free prediction of RMT.
We observe that for small values of $L$, the lattice data are nicely
described by RMT.  We also see that there is a critical scale $L_c$,
the Thouless scale, above which nonuniversal behavior sets in.  In
order to extract this scale from the data, we construct the ratio
\begin{equation}
  \label{eq:ratio}
  \text{ratio}(L)=\frac{\Sigma^2_{\text{latt}}(L)
  -\Sigma^2_{\text{RMT}}(L)}{\Sigma^2_{\text{RMT}}(L)} 
\end{equation}
which should start to deviate strongly from 0 above $L_c$.  We find
that the numerical value of the Thouless scale depends on where we are
in the spectrum, i.e., on the starting point of the interval of length
$L$.  This means that spectral averaging must not be used to construct
$\Sigma^2(L)$ for the purpose of extracting the Thouless energy.

\begin{figure}[!t]
  \begin{center}
    \epsfig{figure=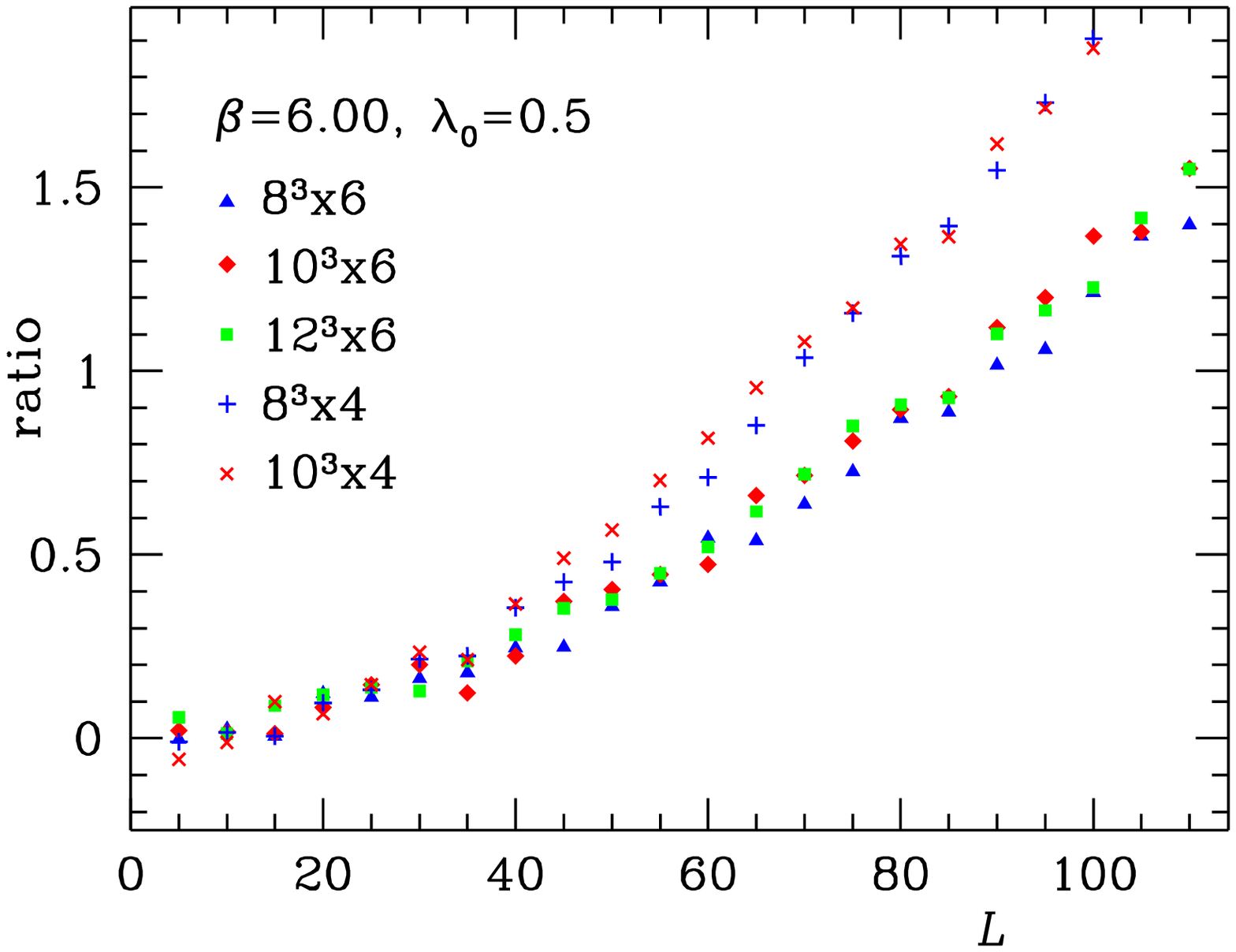,width=65mm}\\[3mm]
    \epsfig{figure=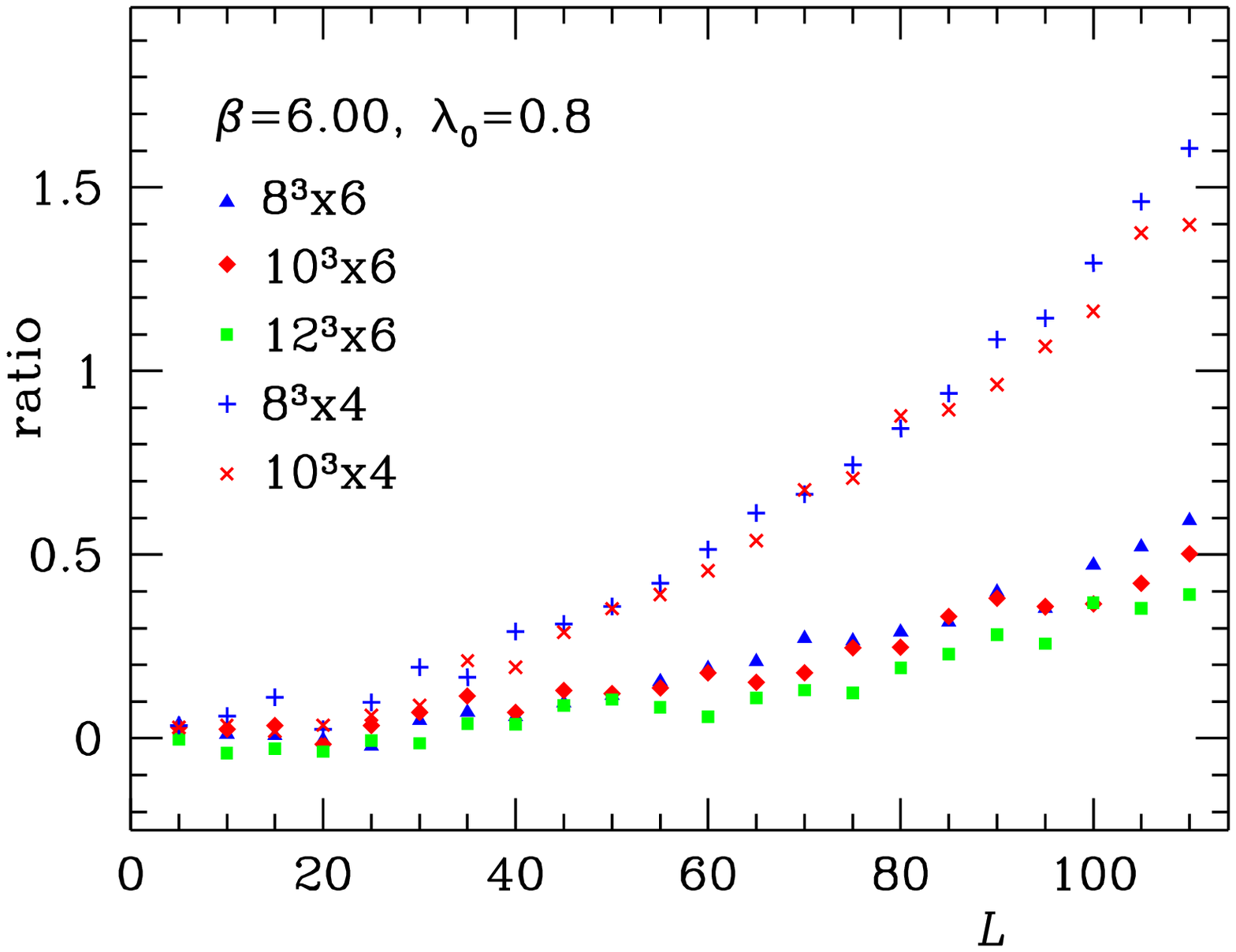,width=65mm}
    \vspace*{-9mm}
    \caption{The ratio defined in Eq.~(\ref{eq:ratio}) for two
      different values of $\lambda_0$ and various lattice sizes.}
    \label{fig:ratio}
    \vspace*{-6mm}
  \end{center}
\end{figure}
In Fig.~\ref{fig:ratio}, we show the ratio of Eq.~(\ref{eq:ratio}) for
two different starting points, $\lambda_0=0.5$ and $\lambda_0=0.8$,
and for various lattice sizes at $\beta=6.0$, which is above $\beta_c$
for both values of $N_t$.  Again, we used 600 independent
configurations per parameter set.  We observe that the Thouless scale
seems to be independent of $N_s$, but depends on $N_t$.  We are
currently investigating the form of this $N_t$ dependence.


\section{LOCALIZATION PROPERTIES OF DIRAC EIGENVECTORS}

In condensed matter physics, the question of whether or not a
disordered mesoscopic sample is a metal or an insulator can be
answered by constructing the so-called inverse participation ratio
$I_2$, which is a measure of how many sites contribute significantly
to the wave function.  For the case of QCD, we introduce the
gauge-invariant definition
\begin{align}
  I_2(\lambda)&\equiv V \frac{\sum_x p_\lambda(x)^2 }{\left[
      \sum_x p_\lambda(x)\right]^2}\:,
  \intertext{where $V$ is the lattice volume and $p_\lambda(x)$ is
      the gauge-invariant probability density} 
  p_\lambda(x) &= \sum_{\alpha = 1}^{N_c} \left|
      \psi^\alpha_\lambda(x) \right|^2\:.
\end{align}
Here, $x$ denotes a lattice site, $\alpha$ is a color index, and
$\psi_\lambda^\alpha(x)$ is a component of the eigenvector
corresponding to eigenvalue $\lambda$.  Because of chiral symmetry we
have $\sum_{x\:\text{even}}p_\lambda(x)=
\sum_{x\:\text{odd}}p_\lambda(x)$.  A completely localized state
therefore has $I_2=V/2$.  This case corresponds to uncorrelated
eigenvalues and is described by the Poisson ensemble. (The
corresponding mesoscopic sample would be an insulator.)  For a
completely delocalized state, $p_\lambda(x)$ is the same for all $x$,
and $I_2=1$.  In RMT, we find
\begin{equation}
  \langle I_2\rangle=\left(1+\frac1{N_c}\right)\frac{N_cV}{N_cV+2}
  \stackrel{V\to\infty}{\longrightarrow}1+\frac1{N_c}\:.
\end{equation}
This case corresponds to extended wave functions with many significant
components  and strongly  correlated eigenvalues.   (The corresponding
mesoscopic sample would  be a metal.)  In QCD,  a simple argument (see
\cite{review}) shows that if chiral symmetry is broken, the Dirac wave
functions must be extended.

\begin{figure}[!t]
  \begin{center}
    \begin{tabular}{c@{\hspace*{5mm}}l}
    \epsfig{figure=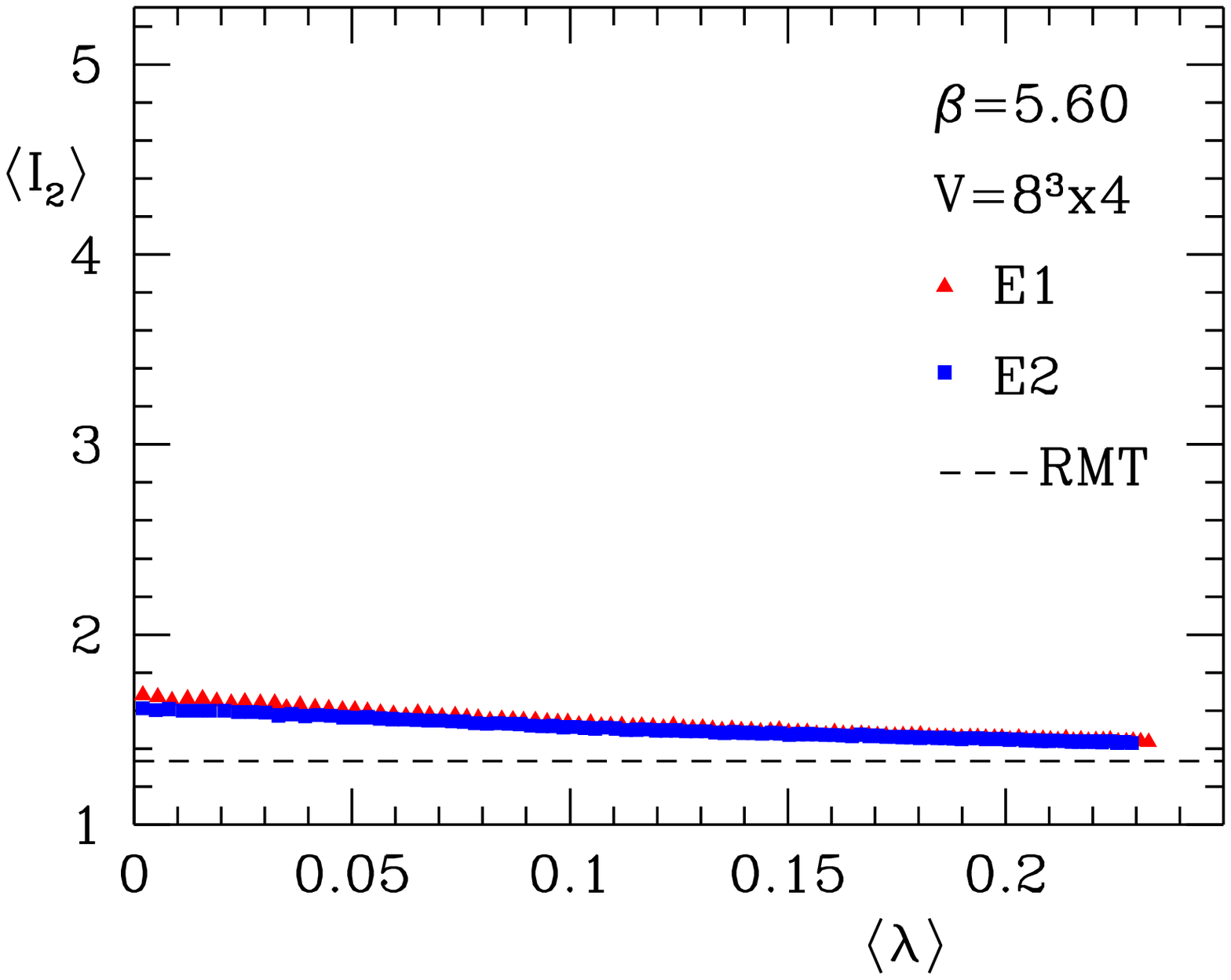,width=49mm} & 
    \raisebox{20mm}{$T<T_c$} \\[2mm]
    \epsfig{figure=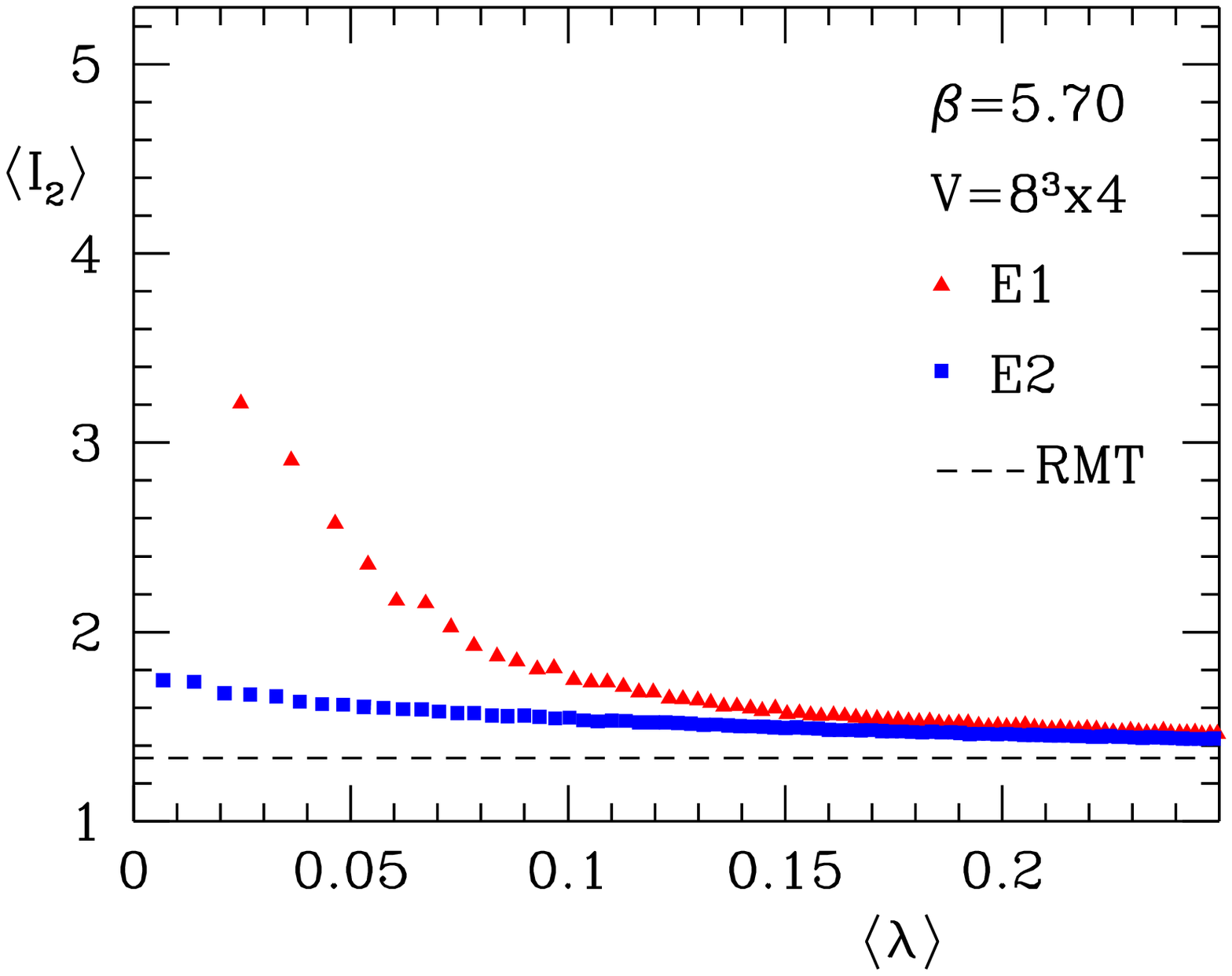,width=49mm} & 
    \raisebox{20mm}{$T\approx T_c$} \\[2mm]
    \epsfig{figure=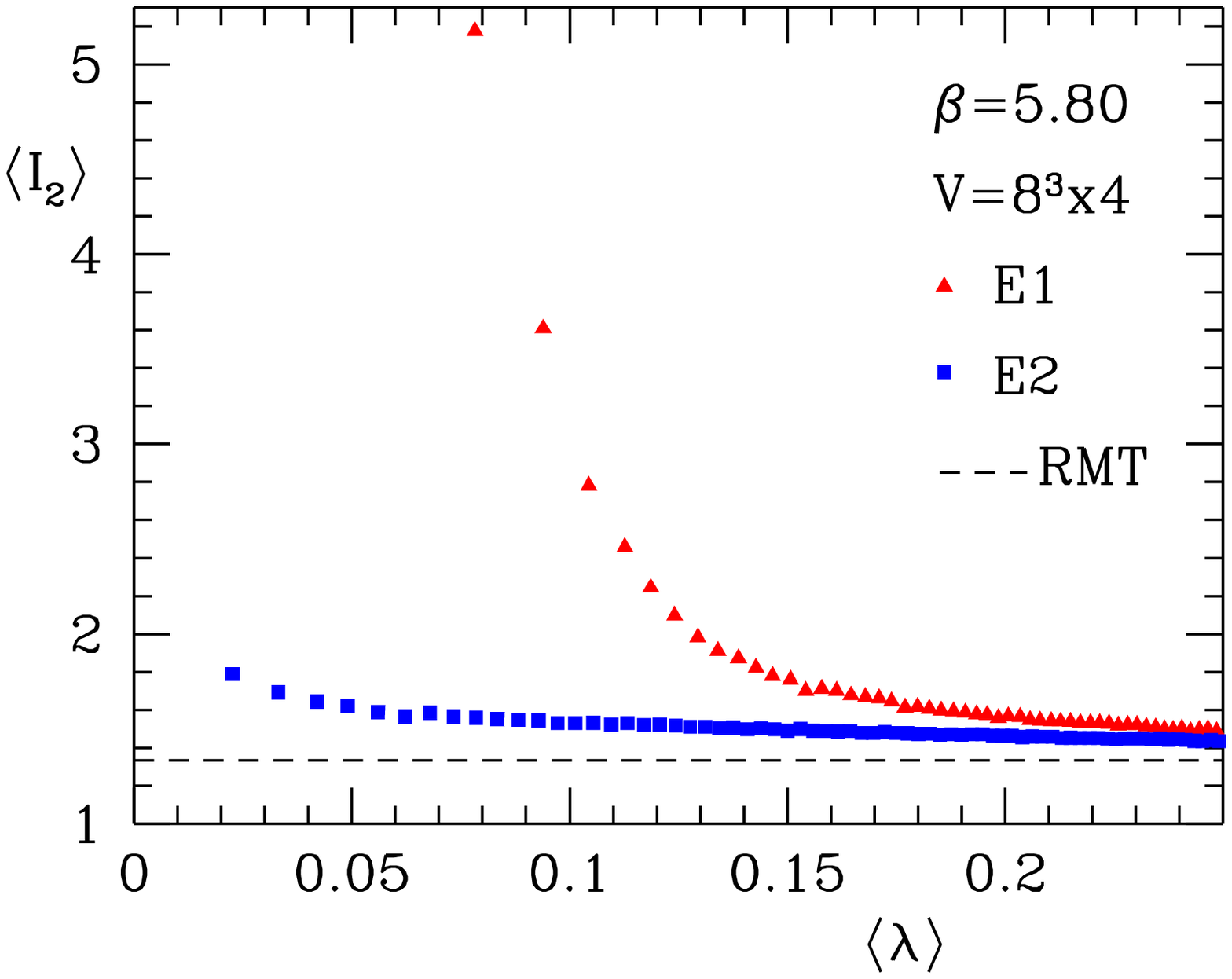,width=49mm} & 
    \raisebox{20mm}{$T>T_c$}
    \end{tabular}
    \vspace*{-8mm}
    \caption{Average inverse participation ratio $\langle I_2\rangle$
      of the low-lying Dirac eigenvectors vs average eigenvalue in the
      different $Z_3$-phases for temperatures below, at, and above
      $T_c$.} 
    \label{fig:IPR}
    \vspace*{-10mm}
  \end{center}
\end{figure}

Our results for the inverse participation ratio of the low-lying Dirac
eigenvectors are shown in Fig.~\ref{fig:IPR}.
Several features are worth noting:\\
1. Below $T_c$ all eigenvectors are extended, which is consistent with
the fact that the eigenvalues are described by RMT both at the hard
edge and in the bulk of the spectrum.  The eigenvectors corresponding
to smaller eigenvalues are slightly more ``localized'' than those
corresponding to larger eigenvalues.  The data agree well with the RMT
prediction, $\langle I_2\rangle=4/3$, for large
$\langle\lambda\rangle$.  In contrast to the observation of
Ref.~\cite{Jans97} for Wilson fermions, we did not find signs of
strong localization.\\
2. The eigenvectors corresponding to eigenvalues in the bulk of the
spectrum remain extended for E1 and E2 at all temperatures considered,
consistent with the fact that the eigenvalues in the bulk
continue to be described by RMT above $T_c$. \\
3. Most interestingly, the eigenvectors of the E1 ensemble
corresponding to the lowest eigenvalues become more localized at and
above $T_c$, while in the E2 sector the lowest eigenmodes remain
extended.

We are currently studying the topological properties of the low-lying
eigenvectors and will pre\-sent our results in the near future.

\section{CONCLUSIONS}

The behavior of the Dirac spectrum at and above the critical
temperature can help to improve our understanding of the chiral phase
transition.  In particular, the properties of the Dirac eigenvectors
deserve further study.  Of course, the quenched approximation and the
use of staggered fermions make it difficult to establish contact with
continuum QCD.  These systematic problems can be circumvented by using
dynamical Ginsparg-Wilson fermions which, however, require large
computational resources.  For the moment, we are restricted to
exploratory studies such as the present one.

\smallskip\noindent{\bf Acknowledgments.} 
This work was supported by DFG project Scha 458/5-3 and by DOE
grants DE-FG02-91ER40608 and DE-AC02-98CH10886.

\end{document}